\documentclass[openacc]{rstransa}

\usepackage{aas_macros}

\titlehead{Review}

\begin{document}

\newcommand{\Om}{\Omega_\mathrm{m}}
\newcommand{\rd}{r_\mathrm{d}}
\newcommand{\lcdm}{$\Lambda$CDM}
\newcommand{\Ob}{\Omega_\mathrm{b}} 
\newcommand{\wcdm}{$w$CDM} 
\newcommand{\wowacdm}{$w_0w_a$CDM} 
\newcommand{\lya}{Ly$\alpha$\space}
\newcommand{\Hrd}{H_0r_\mathrm{d}}
\newcommand{\Mpc}{\, \text{Mpc}}
\newcommand{\kmsMpc}{\,{\rm km\,s^{-1}\,Mpc^{-1}}}
\newcommand{\Planck}{\emph{Planck}}

\newcommand{\dataplus}{\allowbreak+}
\newcommand{\leftparbox}[2]{\parbox{#1}{\begin{flushleft} #2 \end{flushleft}}}
\newcommand{\twoonesig}[4][\pbwidth]{
\begin{equation}
\left.
 \begin{aligned}
#2 \\ #3
 \end{aligned}
\ \right\} \ \ \mbox{\text{\leftparbox{#1}{#4}}}
\end{equation}
}

\title{Future directions in cosmology}

\author{
N. Palanque-Delabrouille$^{1}$}

\address{$^{1}$Lawrence Berkeley National Laboratory, 1 Cyclotron Road, Berkeley, CA 94720, USA}

\subject{cosmology, dark energy}


\corres{Nathalie Palanque-Delabrouille\\
\email{npalanque-delabrouille@lbl.gov}}

\begin{abstract}
Cosmology is entering a very exciting time in its history, when a wealth of cutting-edge experiments are all starting to collect data, or about to. These experiments aim at addressing some of the most intriguing questions in fundamental physics, such as what is the nature of dark matter, is dark energy a cosmological constant or a varying field, what are the masses of the neutrinos, and more. While Lambda-CDM has emerged as a simple model that is consistent with most of the current data sets, we are starting to see some interesting deviations that deserve further exploration. This contribution provides an overview of upcoming projects and the science opportunities they will allow. In particular, we recall and comment the DESI year-1 BAO constraints and their implications for dark energy. We  put some of the most recent results and outstanding questions in the perspective of the forthcoming observational program.
\end{abstract}


\begin{fmtext}

\end{fmtext}


\maketitle
\section{Introduction}
The flat $\Lambda$CDM model has become a reference  for cosmology. With only six free parameters, it  describes with amazing precision a vast number of cosmological observations, none the least being the ground-breaking prediction of  power spectra of the Cosmic Microwave Background (CMB) temperature and polarization anisotropies, as measured  over the whole sky by the \Planck\ experiment~\cite{Planck-2018-cosmology}. 
Despite its  resounding success, the standard cosmological model   calls for further study. While  remarkably simple, it relies on an extraordinary energy budget, consisting of only 5\% of ordinary baryonic matter, 25\% of dark (non-baryonic) matter that is necessary to explain galaxy formation, gravitational lensing observations or the rotation velocity of spiral galaxies, and 70\% of a fluid of unknown nature, dubbed dark energy, and responsible for the accelerated expansion of the Universe that we observe today. In addition, the standard cosmological model assumes an epoch of inflationary expansion in the very early Universe for which there is  no clear evidence yet {  (see \cite{Guth_81, Linde_82} for the original papers that introduced the inflationary paradigm, and \cite{Lyth:1998xn} for a comprehensive review}). It also leaves several key questions unanswered, such as what are the masses of the three neutrino species. 

On the experimental side, measurements of the parameters that describe our  model of the Universe  are becoming increasingly more precise, and tensions have appeared. The two most popular ones are the so-called $H_0$ and $S_8$ tensions, { where $H_0$ is the current value of the expansion rate of the Universe, and $S_8 \equiv \sigma_8\sqrt{\Om/0.3}$ is a measurement of structure growth, with $\Om$ the present-day matter density parameter and $\sigma_8$ quantifying the amplitude of density fluctuations}. The $H_0$ tension shows a 3 to 5~$\sigma$ discrepancy between direct measurements of $H_0$ from a distance-ladder approach,   based in particular on cepheids~\cite{2024ApJ...962L..17R}, and values of $H_0$ derived from fitting  the  $\Lambda$CDM model to CMB data~\cite{Planck-2018-cosmology,Planck-2018-likelihoods}. The $S_8$ tension is a 2 to 3~$\sigma$ discrepancy between CMB and weak-lensing  determinations of  the $S_8$ parameter {  (see  the  recent  weak-lensing results  from the DES~\cite{PhysRevD.105.023520}, KiDS~\cite{Heymans:2020gsg} and  HSC~\cite{Dalal:2023olq}  surveys). Many other  combinations of existing datasets  also indicate various levels of tensions on $S_8$ with the CMB measurements}. 
These tensions could hint to new physics, if not due to unaccounted-for systematics. Recent developments on both tensions are of high interest. Notably, in these proceedings, Wendy Freedman  shows recent $H_0$ measurements from JWST in better agreement with the CMB value,  and Alex Amon suggests that  the $S_8$ discrepancy could be due to an insufficient treatment of the small-scale non-linear regime in weak lensing data. 

Continued testing of the cosmological model is ever-more of interest, as  more opportunities are becoming available, both in terms of experimental techniques (or probes) and in terms of instruments. In this paper, I will start by giving in section~\ref{sec:probes} an overview of the different methods of relevance to make progress on our understanding of the dark Universe. Cosmology is entering a very exciting time, with a wealth of cutting-edge experiments  all starting to collect data, or about to. In section~\ref{sec:exps}, I will briefly describe these new instruments and their connection to the afore-mentioned observational programs. I will then give a high-level summary of the recent results from the Dark Energy Spectroscopic Instrument (DESI) in section~\ref{sec-DESI}, and I will conclude in section~\ref{sec:future} with an outlook on prospects for future cosmology programs. 

\section{Probes to test our cosmological model}\label{sec:probes}

As was already identified in the  Dark Energy Task Force report~\cite{Albrecht:2006um}, major advances on our understanding of dark energy are expected from a combination of three probes: Type Ia supernovae, Baryon Acoustic Oscillations and Weak Lensing. We briefly introduce these three probes below. {  Other approaches are sometimes also considered but haven't quite yet reached the same level of robustness. Among these, a common alternative approach to test cosmology is for instance the dependence with redshift of cluster abundance.}

\subsection{Type Ia supernovae (SNIa)}
The accelerated expansion of our Universe was first discovered at the end of the XX$^{\rm th}$ century using Type Ia supernovae, or SNIa. These stellar explosions {  occur in binary systems where a white dwarf accretes matter from its stellar companion, until it reaches a critical mass of about $1.4~M_\odot$ beyond which the electron degeneracy pressure can no longer counter-balance the gravitational pressure, unbinding the star in a thermonuclear supernova explosion. Type Ia supernovae are  cosmological candles that can be standardized to high precision, which  allows them to be excellent cosmological probes. }
To this day, SNIa  continue to be one of the main probes to study the expansion of the Universe {  through the distance-redshift relation}.  The Hubble diagram representing the apparent magnitude of distant SNIa as a function of redshift can typically be measured to redshifts of order $z=2$. It is used to determine the dark matter and dark energy composition of the Universe. Several experiments are dedicated to the discovery and follow-up of SNIa for this purpose. The Pantheon+ compilation~\cite{Scolnic:2021amr}  includes 1550 spectroscopically-confirmed SNIa to $z=2.26$.  The Union3 compilation~\cite{Rubin:2023} includes 2087 spectroscopically-confirmed SNIa, of which about 1360 are in common with the Pantheon+ sample. Using their 5-year data set, the DES project released a homogeneous selection of 1635 photometrically-classified SNIa~\cite{DES:2024tys}, mostly independent from the previous two samples in the distant Universe, but sharing with both of them about 194 historical SNIa at $z<0.1$. 

\subsection{Large-scale clustering: Baryon Acoustic Oscillations (BAO) and Redshift-Space Distortions (RSD)}
Baryon Acoustic Oscillations (BAO) are imprints of the propagation of plasma waves in the early Universe, frozen at the epoch of recombination ($z\sim 1100$) when the temperature of the Universe dropped sufficiently for the first atoms to form and the Universe to become non-ionized. The distance these sound waves propagated until recombination can be used as a standard ruler, which only evolves thereafter under the expansion of the Universe. Its dimension at different epochs is therefore a tracer of the expansion of the Universe. It is a 3D measurement. The observed dimension across the line of sight, $\Delta\theta$ is directly related to the angular diameter distance $D_M(z)$ through $\Delta\theta = r_d/D_M(z)$, while the observed dimension along the line of sight, $\Delta z$, provides a measurement of the Hubble parameter $H(z)$ via $\Delta z = H(z)r_d/c$, where $r_d$ is the sound horizon at the baryon drag epoch of the early Universe, of comoving size $r_d\sim 150$~Mpc, which here serves as a standard ruler. 

In addition to BAO,  the clustering of galaxies can be used to study the overall properties of gravity via the Redshift-Space Distortion (RSD) effect: an apparent flattening in redshift-space of the spherically symmetric clustering of matter in real space, due to the linear infall of matter onto gravitation wells. The apparent redshift of a galaxy is the sum of its cosmological redshift and a contribution from its peculiar velocity  along the line of sight. The Kaiser formula gives the observed clustering power spectrum $P(z,k,\mu)$ at redshift $z$ and on scale $k$ as 
\begin{equation*}
P(z,k,\mu) = \left[b+f\mu^2\right]^2P_m(k)
\end{equation*}
where $b$ is the galaxy bias, $f$ the linear growth rate and $\mu$ the cosine of the angle of the galaxy pair with respect to the line of sight, and $P_m(k)$ is the linear power spectrum of matter. Since $P_m(k)\propto\sigma_8^2$, the RSD effect therefore allows us to constrain $f\sigma_8$, and, through $f$, to study gravity and test General Relativity. 

\subsection{Weak lensing (WL)}
The third approach commonly used to study dark energy is weak lensing. It  uses the distortion of background images due to the bending of light in the vicinity of galaxies or clusters of galaxies. In practice, weak lensing experiments measure the power spectrum of the galaxy shape distortion signal. Weak lensing is directly probing the distribution of mass. Recently, weak lensing has often been combined in a set of 3 measurements of 2-point correlation functions or power spectra: weak lensing (or shape auto-correlation), galaxy auto-correlation, and  cross-correlation of shape and galaxy distributions (also called galaxy-galaxy lensing). These 3x 2-point studies are typically aimed at constraining $\sigma_8$ or $S_8\equiv\sigma_8\times({\Om/0.3})^{0.5}$.

Interestingly, galaxy-galaxy lensing and RSD can be used together to directly measure the linear growth rate $f$, and thus constrain general relativity. This is done through the ratio $E_G$ defined as 
\begin{equation*}
E_G \equiv \frac{1}{\beta}\frac{\Upsilon_{gm}}{\Upsilon_{gg}}
\end{equation*}
where $\beta=f/b$, and $\Upsilon_{gg}$ and $\Upsilon_{gm}$ are the galaxy-galaxy and galaxy-matter differential surface densities, respectively~\cite{Reyes2010-EG}. Because the galaxy bias $b$ enters squared in the former and linearly in the latter, both $b$ and $\sigma_8$ terms cancel out in the $E_G$ statistics, leaving $f$ as the only free parameter. This is an approach that could be developed more in the future with the advent of new instruments dedicated to each of these observational techniques. 

\section{A new generation of experiments to study Dark Energy} \label{sec:exps}
Beginning 2020, the Stage-IV generation of Dark Energy experiments are one by one coming online. These projects present interesting complementarities in terms of methodology, redshift coverage, and instrumental properties, which will allow them to probe the dark Universe from different perspectives. Table~\ref{tab:exps} summarizes the main parameters of these projects. The first one to have started taking data is DESI, the Dark Energy Spectroscopic Instrument, and we will devote section~\ref{sec-DESI} to the presentation of its first-year BAO results.  
\begin{table}[!h]
\centering
\caption{Summary of forthcoming Dark Energy experiments. In the Probes column, WL stands for weak lensing, LSS for large-scale structure and SNIa for Type Ia supernovae.}
\label{tab:exps}
\begin{tabular}{llllll}
\hline
Name & Probes & Telescope & Survey  & Start  &  Survey \\
& & size&area (deg$^2$) & date & duration \\
\hline
DESI & LSS & 4m& 14,000 & May 2021 & 5 years \\
Euclid & WL \& LSS & 1.2m & 15,000 & Feb. 2024 & 5 years \\
PFS & LSS & 8.2m & 1,400  & mid-2024 & 100 nights \\
Rubin & WL \& SNIa & 8.4m & 18,000  & end 2025 (first light) & 10 years \\
Roman & WL \& LSS \& SNIa & 2.4m & 2,000  & 2027 (launch) & 5 years\\
\hline
\end{tabular}
\vspace*{-4pt}
\end{table}

The Euclid space mission~\cite{2022A&A...662A.112E} is the flagship project of the European Space Agency. Equipped with a wide-field instrument in the visible (a large band roughly covering  $r$+$i$+$z$), it will observe over 100~million galaxies for weak-lensing studies. Redshifts for these galaxies will be determined from photometry. The main advantage of Euclid is to be located in space, while all  previous  WL experiments  were ground-based. This provides it with a significantly better image resolution, a key asset for the shape measurements required in WL studies, thanks to a diffraction-limited seeing of $0.16"$. 
Euclid is also equipped with slit-less spectroscopy (resolution $R\sim 450$ in the near IR), which will allow it to measure the redshifts of 25 million galaxies (from a selection of 1700 H$\alpha$ emitters per deg$^2$) in the redshift range $0.9<z<1.8$. These will be used for BAO and RSD studies. 

The Subaru Prime Focus Spectrograph (PFS)~\cite{2014PASJ...66R...1T} is a massively-multiplexed fiber-fed optical and near-infrared 3-arm spectrograph, equipped with 2400 fibers. The instrument has a good resolution $R\sim 4500$ in the red and near IR, allowing it to measure the spectra of 1~million emission-line galaxies in the redshift range $0.8<z<2.4$. The goal of the project is to reach an uncertainty of $3\%$ on the expansion rate $H(z)$ in each of 6 redshift bins. Compared to Euclid or DESI, PFS covers slightly higher redshifts. Unlike DESI or Euclid, however,  the wide-field spectroscopic survey of PFS is not a dedicated cosmology survey. The cosmology survey will be conducted over a total period of 100~nights, during which it will observe a footprint of 1,200~deg$^2$,  an order of magnitude smaller than the sky coverage of DESI or Euclid.

The flagship ground-based project for WL studies is the Vera Rubin (previously LSST) observatory~\cite{2019ApJ...873..111I}, in particular its DESC collaboration. Rubin will conduct a WL survey with about 2 billion galaxies observed in each of the standard visible ($u$, $g$, $r$, $i$, $z$) and near IR ($Y$) bands. Using extremely short exposures of 15~sec only, Rubin will be able to  reduce the impact of atmospheric perturbations on the image quality, a key parameter for the resolution of galaxy shapes, and expects to achieve a median seeing of 0.7 arcsec. While this is excellent for a ground-based instrument, it does not compete with the $0.16$~arcsec seeing achievable with Euclid from space.  Rubin, however, presents the advantage of a longer survey duration and a much larger primary mirror than Euclid, hence allowing for much deeper and less noisy observations. { Another key advantage of LSST over Euclid is its ability to obtain photometric redshifts for all its sources given its multi-band capabilities.}  In addition to WL, Rubin will also lead a survey of millions of SNIa to redshift $z<0.5$, roughly 10 times larger than all existing SNIa surveys combined.  This will provide unique SNIa statistics, allowing precise studies of dependence on SNIa populations.

The Nancy Roman space telescope~\cite{Wang_2022}, a project of the NASA, will conduct a high-latitude spectroscopic survey covering 2000~deg$^2$. Using a wide-field imager in the visible and near IR, it will measure the shape of galaxies for WL studies. With slitless spectroscopy (resolution $R\sim 460$  in the near IR), it will measure redshifts of about 10 million H$\alpha$ galaxies in the redshift range $1<z<2$ and about 2 million [OIII] galaxies in the redshift range $2<z<3$, both for large-scale structure studies via BAO and RSD. Roman is focusing on higher redshifts than DESI or Euclid, and with higher densities of galaxies in the overlapping redshift region, but covers a much smaller area. Roman also plans a 6-month SNIa survey  over 2 years to observe about 12,000 SNIa in the redshift range $0.5<z<2.0$, making it a unique project aiming at the study of dark energy with all three standard probes. Although the project is nominally planned for 5 years, a 5-year extension is considered. 

The first Stage-IV dark energy experiment that came online is the Dark Energy Spectroscopic Instrument DESI~\cite{DESI2016a.Science,DESI2022.KP1.Instr}. DESI is designed to  significantly improve our current knowledge on dark energy through the study of large-scale structure (BAO and RSD).  DESI has started observing over 40 million galaxies and quasars over a continuous coverage of the redshift range from 0 to above 4. The focal plane contains 5000 robotic fiber positioners, the largest currently operating or  in construction, guiding the light collected in the focal plane to 10 3-arm spectrographs covering the entire visible range from 360~nm to 980~nm. DESI has been taking survey data since May 2021, and released its first year BAO results in April 2024. These results are presented in the following section. The nominal DESI survey is planned for 5 years, and the collaboration recently presented on a 2.5-year extension to run until December 2028.

\section{DESI year-1 BAO results}\label{sec-DESI}

\subsection{DESI year-1 data}
To optimally cover its full redshift range, DESI uses a series of tracers:  13 million sources of the Bright Galaxy Survey (BGS) up to redshift $z<0.4$,  8 million luminous red galaxies (LRG) to cover the redshift range $0.4<z<1.0$,  16 million young blue emission-line galaxies  (ELG) at  redshifts $0.6<z<1.6$ and finally 3 million quasars in the most distant Universe, both as direct tracers in the range $1.0<z<2.1$ and as background sources to trace the intergalactic distribution of neutral hydrogen at redshifts $z>2.1$. 

The year-1 DESI sample consists of data taken between May 2021 and July 2022. It includes 5.7 million unique galaxies and  quasars up to redshift 2, and 420,000 \lya forest lines of sight. Collected in a single year of observation, this sample  already consists of more than twice the 2.8 million redshifts used in the final SDSS cosmological analysis and collected over two decades of SDSS observation~\cite{2021PhRvD.103h3533A}. The year-1 data cover a footprint of roughly 7,500~deg$^2$. Because the various target classes are observed at different priorities, with the rare quasars at highest priority and the abundant ELGs at lowest priority, this sample contains 45\% of the quasars expected to be observed in the full survey, and only about 25\% of the ELGs. The later observations will complete the sample to achieve the forecasted densities for all classes.  

Thanks to the high statistics of this first data set, we can divide the LRG sample in 3 independent redshift bins ($0.4-0.6$, $0.6-0.8$ and $0.8-1.1$), and the ELG sample into 2 independent redshift bins ($0.8-1.1$ and $1.1-1.6$). We combine the highest LRG and lowest ELG bins that cover the same redshift range into a common sample, and thus end up with a total of  6 independent measurements of the 2-point correlation function across the full redshift  range. The combined LRG+ELG bin alone leads to the detection of the BAO feature at an effective redshift of 0.93 with a  significance of 9$\sigma$ and a precision on the distance-scale measurement of 0.8\%. Overall,  the full sample achieves an aggregated precision of $0.49\%$ from this single year of observation~\cite{DESI2024.III.KP4}, which compares favorably with the $0.60\%$ precision for the final results from 20 years of SDSS observations. A thorough study of systematic uncertainties showed that all effects combined led to an impact smaller than 0.3\% on the determination of the BAO distance-scale, {  with largest contributions  coming from the galaxy-halo connection, the theoretical modeling of BAO, and the fiducial cosmology assumed when converting angles and redshifts to distances.} The systematic uncertainty  only increased the total uncertainty by 5\% when added in quadrature with the statistical uncertainty. 

The so-called \lya forest that is observed in the spectra of distant quasars is caused by the absorption of the background quasar light by the neutral hydrogen along the line of sight, mostly occurring at the wavelength of the \lya transition.  This results in a series of absorption features that compose the  \lya forest and can be used to probe the matter distribution in the high-redshift Universe (at $z>2.1$ for the absorption feature to fall in the visible band). The \lya forest data are analyzed separately from the galaxies and quasars, as in this case the cosmological information of interest for the BAO measurement is not solely the redshift of the source but is embedded in the full spectrum of the observed quasar. These data are used in two ways: either as autocorrelations of \lya absorptions across two different lines of sight, or as cross-correlations between \lya forest pixels and quasars at similar redshifts.  The DESI \lya sample achieves a precision of 1.1\% on the measurement of the BAO feature at an effective redshift $z=2.33$~\cite{DESI2024.IV.KP6}, with a systematic uncertainty of 0.5\%.

The BAO feature shows as a peak in the 2D autocorrelation function of galaxy, quasar or \lya  forest distribution. The location of that peak contains  the cosmological information of interest to study the expansion history of the Universe and therefore dark energy. 

\subsection{Cosmological interpretation of DESI year-1 results}

\begin{figure}[!h]
    \centering
    \includegraphics[width = \textwidth]{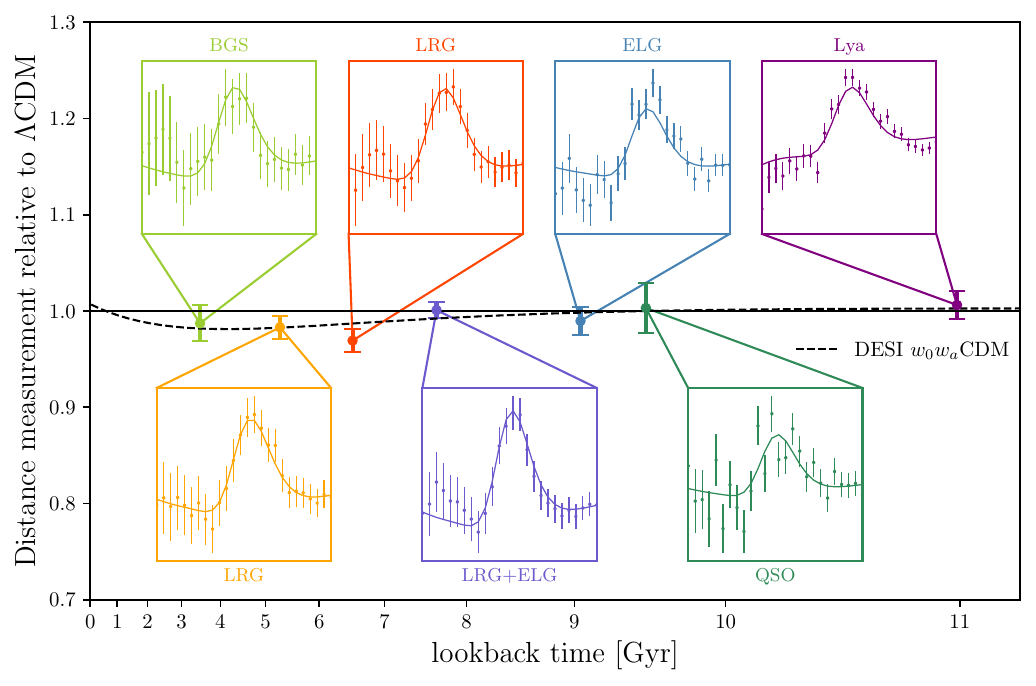} 
    \caption{Hubble diagram of the BAO distance scale. The insets zoom in on the BAO region of the  isotropic correlation function for each of the tracer samples. The reference line at unity is the best-fit \lcdm\ model. The dashed line is the best-fit  \wowacdm\ model. Credit: Arnaud de Mattia / DESI collaboration.}
    \label{fig:BAO_DV}
\end{figure}

All the cosmological results summarized in this section are taken from~\cite{DESI2024.VI.KP7A}. The observed size of the BAO feature across and along the line of sight translates into measurements of $D_M/r_d$ and $D_H/r_d$ respectively. The distances $D_M$ and $D_H$ can be combined into an isotropic distance scale measurement $D_V$ according to
\begin{equation}
D_V = \left[zD_M(z)^2 D_H(z)\right]^{1/3}.
\end{equation}
For low-statistics data, the BAO measurement cannot be decomposed into an independent determination of the distance scale in 3D and only the isotropic measurement can be made. With the year-1 sample, this is the case for the BGS and the quasar data. Figure~\ref{fig:BAO_DV} shows the summary, in the form of a Hubble diagram, of the BAO distance-scale $D_V$ as a function of redshift, in the context of \lcdm. DESI year-1 results are in good agreement with \lcdm\ (flat solid line centered on 1).

\begin{figure}[!h]
    \centering
    \includegraphics[width = 3.5in]{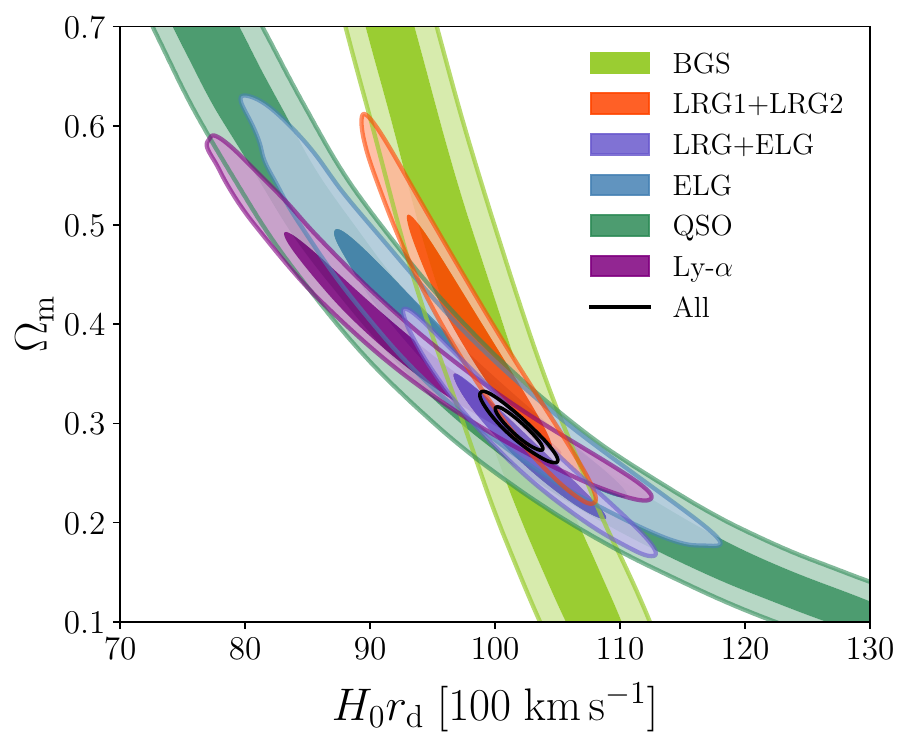} 
    \caption{68\% and 95\% credible-interval contours for parameters $\Om$ and $\Hrd$ obtained for a flat \lcdm\ model from fits to BAO measurements from each DESI tracer type individually, as labeled (from~\cite{DESI2024.VI.KP7A}). Results from all tracers are consistent with each other and the change in the degeneracy directions arises from the different effective redshifts of the samples.}
    \label{fig:Om_H0_rd}
\end{figure}

The BAO distance-scale measurements constrain only two free parameters, which provide measurements of  the matter density $\Om$ and the product $\Hrd$ of the Hubble constant $H_0$ and the sound horizon $\rd$. Figure~\ref{fig:Om_H0_rd} shows the 68\% and 95\% credible-interval contours in this plane for a \lcdm\  model from each of the tracers.  It is worth noting that the axis of the degeneracy evolves with redshift, making the different tracers sensitive to a different combination of $\Om$ and  $\Hrd$. As seen in the figure, the individual measurements are in excellent agreement with each other, allowing us to combine them and achieve a 5.1\% measurement of $\Om$ and a 1.3\% measurement of $\Hrd$ from DESI alone:
\twoonesig[1.5cm]
{\Om &= 0.295\pm 0.015,}
{\rd h &= (101.8\pm 1.3) \Mpc,}
{DESI~BAO, \label{eq:DESI_LCDM_results}} 
where for the sake of simplicity  we quote results for $\rd h\equiv\Hrd/(100\;\kmsMpc)$ instead of $\Hrd$.

\begin{figure}[!h]
    \centering
    \includegraphics[width = 0.8\textwidth]{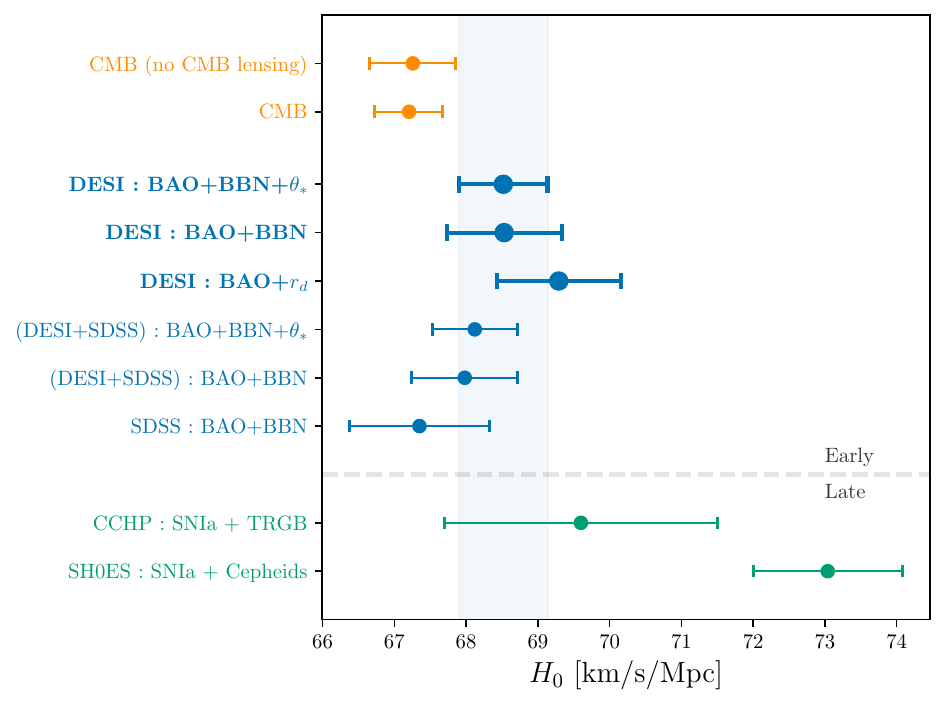} 
    \caption{$68\%$ credible-interval constraints on the Hubble constant, assuming the flat \lcdm\ model (from~\cite{DESI2024.VI.KP7A}). The blue, bold whiskers show DESI BAO measurements in combination with an external BBN prior on $\Ob h^2$ and measurement of the acoustic angular scale $\theta_\ast$, or the BBN prior alone. The thin blue whiskers show the corresponding results from SDSS BAO.
        The orange whiskers show the results from CMB anisotropy measurements from \Planck\ and ACT, while the green whiskers show measurements of $H_0$ from the distance ladder with either Cepheids or TRGB.  }
    \label{fig:H0}
\end{figure}

While BAO alone only constrains the product $\Hrd$, additional data can break the degeneracy by providing information to determine $\rd$. Notably, Big Bang Nucleosynthesis results constrain $\Ob h^2$ and therefore are sufficient to determine $\rd$~\cite{Addison:2013, 2015PhRvD..92l3516A}. Within flat \lcdm, combining DESI with BBN yields a 1.2\% determination of $H_0$:
\begin{equation}
\label{eq:H0_DESI-bbn}
H_0= (68.53\pm 0.80)\kmsMpc
\qquad
(\mbox{DESI~BAO\,+\,BBN}).    
\end{equation}
The DESI year-1 BAO results on $\Om$ and $H_0$ are consistent with CMB. In particular, the measurement of $H_0$ is in agreement with CMB results and in tension at the 3.7$\sigma$ level with  the SH0ES Cepheid-based distance-ladder result   $H_0=73.04\pm 1.04\;\kmsMpc$ of  \cite{Riess:2021jrx}  (see figure~\ref{fig:H0}).

The main objective of DESI is to study dark energy. As a first step, we considered a \lcdm\ model and tested the implications of DESI year-1 results against it. In a second stage, we focused on any departure from the simplest \lcdm\ model and explored models allowing dark energy to be described by a time-varying equation of state. 

The physics of the acceleration of the expansion rate of the Universe can be treated as an effective dark energy density and pressure. The dark energy equation of state (EoS) is the ratio $w$ of the pressure to the energy density. A simple cosmological constant $\Lambda$ would be described by a constant EoS parameter $w=-1$. Therefore, any deviation from $w=-1$ would be a sign that a \lcdm\ model does not capture the dynamics of the background cosmology. The converse however is not true. This is known as the ``mirage of $\Lambda$''~\cite{Linder2007} ({  see~\cite{Akarsu:2023mfb, Akarsu:2024eoo} for  a non-\lcdm\ model that nevertheless has $w=-1$ except during possibly instantaneous transitions}). 

\begin{figure}
    \centering
    \includegraphics[width = 0.68\columnwidth]{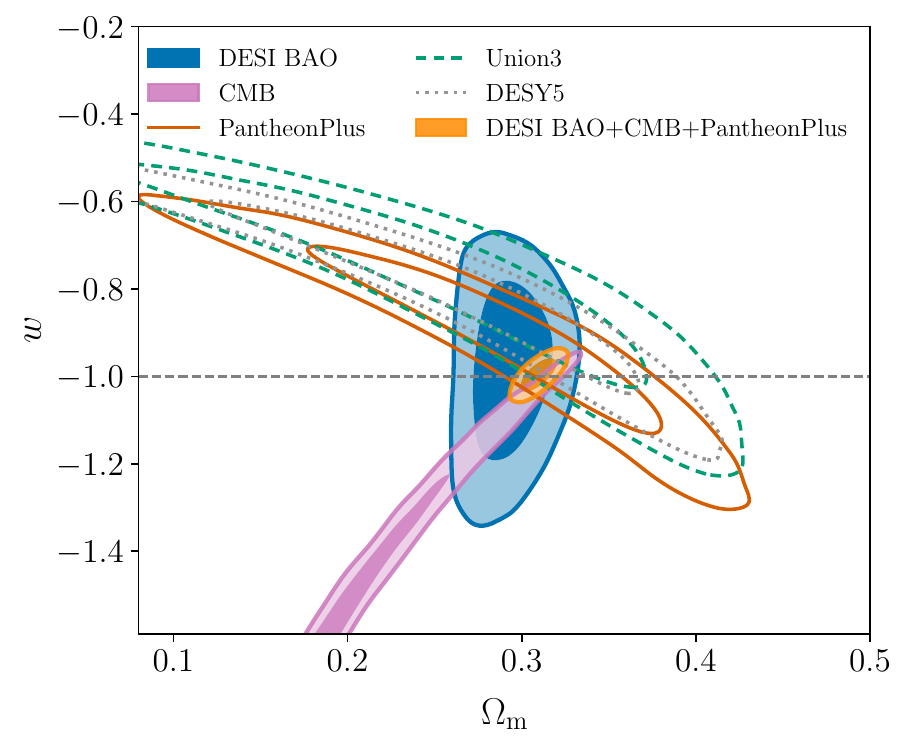}
    \caption{Constraints on $\Om$ and $w$ in the flat \wcdm\ model (from~\cite{DESI2024.VI.KP7A}). The constraints from DESI BAO alone are shown in blue, those from the CMB in pink, and different SN~Ia compilations in solid and dashed green. The orange contour shows the combined constraint from DESI, CMB and PantheonPlus SN~Ia. All contours show 68\% and 95\% credible intervals. Note the remarkable complementarity of cosmological probes in this plane.}
    \label{fig:omegam_w}
\end{figure}

We show in Figure~\ref{fig:omegam_w} results on $w$ and $\Om$ from several probes: CMB, DESI BAO, and various SNIa datasets. Here, we use the shorthand notation ``CMB'' to denote the combination of temperature and polarization data from \Planck, and CMB lensing data from the combination of \Planck\ and ACT~\cite{Carron:2022,Madhavacheril:ACT-DR6}. All these probes show good agreement with $w=-1$, and their combination yields a better than 3\% measurement of $w$ that is consistent with a \lcdm\ model:
\twoonesig[3.5cm]
{\Om &= 0.3095\pm 0.0069,}
{w   &= -0.997\pm 0.025,}
{DESI+CMB\dataplus PantheonPlus. \label{eq:DESI_SN_CMB_wCDM}}

\begin{figure}
    \centering
    \includegraphics[width = 0.48\columnwidth]{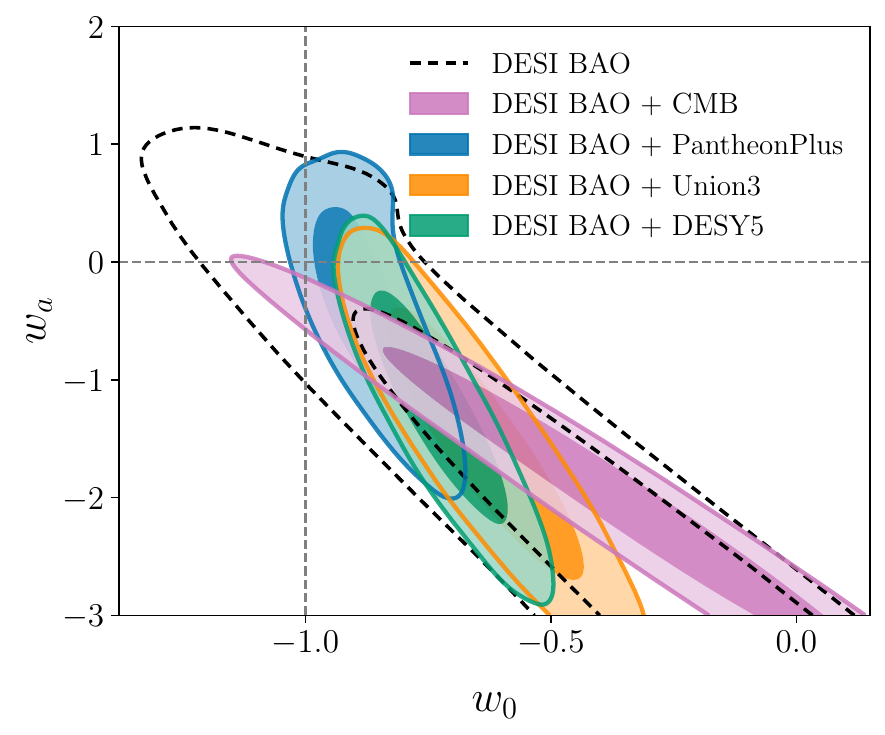} 
    \includegraphics[width = 0.47\columnwidth]{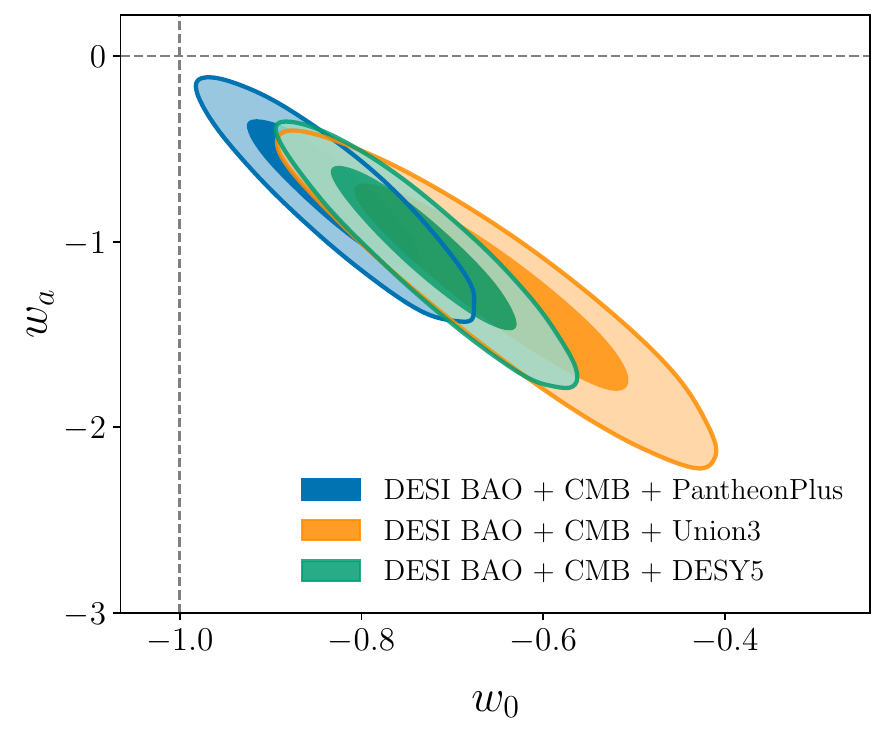}
    \caption{\emph{Left panel}: 68\% and 95\% marginalized posterior constraints in the $w_0$--$w_a$ plane for the flat $w_0w_a$CDM model, from DESI BAO alone (black dashed), DESI + CMB (pink), and DESI + SN~Ia, for the PantheonPlus \cite{Brout:2022}, Union3 \cite{Rubin:2023} and DESY5 \cite{DES:2024tys} SNIa datasets in blue, orange and green respectively. Each of these combinations favors $w_0>-1$, $w_a<0$, with several of them exhibiting mild discrepancies with \lcdm\ at the $\gtrsim2\sigma$ level. However, the full constraining power is not realized without combining all three probes. \emph{Right panel}: the 68\% and 95\% marginalized posterior constraints from DESI BAO combined with CMB and each of the PantheonPlus, Union3 and DESY5 SN~Ia datasets. The significance of the tension with \lcdm\ ($w_0=-1$, $w_a=0$) estimated from the $\Delta \chi_\mathrm{MAP}^{2}$ values is $2.5\sigma$, $3.5\sigma$ and $3.9\sigma$ for these three cases respectively. (from~\cite{DESI2024.VI.KP7A}))}
    \label{fig:w0wa_1}
\end{figure}

Extending to a model with time-varying EoS, we use  the CPL parametrization $w(a) = w_0 + w_a(1-a)$ that has been demonstrated to match the background evolution over a wide range of physically-motivated scenarios~\cite{Linder2003,dePutter2008}.  We constrain the $w_0-w_a$ parameter space using BAO data in combination with other data sets. The previous datasets (DESI BAO, CMB and SNIa) again show good consistency in the $w_0-w_a$ plane, allowing them to be combined. The year-1 DESI BAO data together with CMB data yield best fitting parameters that show a preference over a \lcdm\ model at the 2.6~$\sigma$ level. SNIa data alone also indicate a preference for an evolving EoS with $w_a<0$. Combining DESI and CMB data with any of the three   recent SNIa samples shows a preference for a dark energy EoS described by $w_0>-1$ and $w_a<0$, and deviates from a simple EoS with constant $w=-1$ (as expected for \lcdm)  at a level that varies between $2.5~\sigma$ and $3.9~\sigma$ depending on the choice of SNIa sample used in the data combination. The {  results} for the various combinations are summarized in table~\ref{tab:cosmo} and illustrated in Figure~\ref{fig:w0wa_1}. The best-fit \wowacdm\ model is illustrated as the dashed line in Figure~\ref{fig:Om_H0_rd}.  

\begin{table*}
\centering
\caption{
\label{tab:cosmo}
Cosmological parameter results in a model that allows for time-evolving EoS for dark energy using DESI year-1 BAO data in combination with external datasets.    Results are quoted for the marginalized means and 68\% intervals in each case.  
}
\scriptsize
\begin{tabular}{l c c c c c}
\hline\hline
Dataset  & $\Om$ & $H_0$  & $w_0$ & $w_a$  & Deviation  \\
& &($\kmsMpc$)& & &from \lcdm \\
\hline
    DESI+CMB & $0.344^{+0.032}_{-0.027}$ & $64.7^{+2.2}_{-3.3}$ &  $-0.45^{+0.34}_{-0.21}$ & $-1.79^{+0.48}_{-1.0}$ & 2.6~$\sigma$  \\
    DESI+CMB+Panth. & $0.3085\pm 0.0068$ & $68.03\pm 0.72$ &  $-0.827\pm 0.063$ & $-0.75^{+0.29}_{-0.25}$ & 2.5~$\sigma$  \\
    DESI+CMB+Union3 & $0.3230\pm 0.0095$ & $66.53\pm 0.94$ &  $-0.65\pm 0.10$ & $-1.27^{+0.40}_{-0.34}$ & 3.5~$\sigma$   \\
    DESI+CMB+DESY5 & $0.3160\pm 0.0065$ & $67.24\pm 0.66$ &  $-0.727\pm 0.067$ & $-1.05^{+0.31}_{-0.27}$ & 3.9~$\sigma$ \\
\hline
\end{tabular}
\end{table*}

{  Dark energy parameterization through the EoS, however, remains limited in capturing some aspects of dark energy dynamics. The results presented above hint at models that are phantom (i.e. $w<-1$) for most  of the evolution. This can be an artefact of the CPL approach, and other parameterizations do not show the same behavior. For example, a Chebyshev polynomial expansion of $w(z)$ hint towards an evolving and emergent dark energy, with negligible presence at $z>1$~\cite{DESI:2024aqx}.}

{  The intriguing evidence discussed above for a time-varying EoS deserves careful consideration. An interesting test for potential systematics in the DESI measurements is the comparison with prior SDSS results. The latter are more constraining than the DESI year-1 results up to $z_{\rm eff}=0.6$. Replacing the lowest three DESI redshift bins (the BGS and two lowest-redshift LRG bins) with the  $z_{\rm eff}=0.15$, 0.38 and 0.51 SDSS data points yields fully compatible results on $\Om$ and $\rd h$ with only about 20\% smaller errors, and the results on $w_0,\,w_a$ are in excellent agreement as well.  This indicates that the low-redshift DESI measurements are not significantly driving the final dark energy results. There is also an about 2.5~$\sigma$ discrepancy between the DESI $z_{\rm eff}=0.71$ and the SDSS $z_{\rm eff}=0.70$ values of $D_V$, but we have not identified any source of non-statistical discrepancy that could explain it. We reiterate the value of the blinded procedure performed for the DESI analysis, and we look forward to the DESI year-3 results to get further understanding of the nature of dark energy. 
}

\vspace*{-5pt}

\section{Future perspective in cosmology}\label{sec:future}
The current epoch is unique in the history of cosmology. In the next few years, several experiments dedicated to the study of dark energy will start taking data and address the question of its nature from different perspectives. The DESI project has just released the BAO results from the first year of survey data, but many more results are expected to come. In particular, future publications from DESI, Euclid and LSST will also provide information on the growth of structure and address the current tensions on $\sigma_8$ or $S_8$, will refine the cosmological measurements on neutrino masses that are currently limited to upper bounds, and will continue to probe the dark Universe and test the tantalizing hint for time-varying dark energy. In the next few years, additional data  will also probe the high-redshift Universe and allow us to study the primordial inflationary era, in a complementary approach to that provided by CMB data. 

There are two regimes that are yet poorly explored and where a future generation of spectroscopic surveys could have a major role to play.  Clustering data focusing on the $z>2$ redshift range will constrain the expansion and gravitational growth history in the matter-dominated regime. Probing a larger volume than current surveys, and at redshifts where non-linear growth is limited to smaller scales than at later times, these high-redshift data are well correlated with initial conditions. They are in particular well suited to probe primordial inflation  through a test of the existence of primordial non-Gaussian features such as described by an $f_{\rm NL}$ parameter. With significantly increased target densities at $z<2$, a next-generation survey can further explore the hints of tension with a \lcdm\ model that recently appeared in current surveys. Such a stage-5 spectroscopic instrument will require technical upgrades compared to DESI. Several projects are already being discussed, such as the Wide Spectroscopic Telescope in Europe, or the Spec-S5 project in the US.

\section{Conclusion}
The \lcdm\ model is an amazing fit to almost all cosmological observables from $z=0$ to $z\sim1000$. Yet, tensions have appeared between results obtained with CMB data and those from other approaches, which need to be resolved. This is in particular the case for the determination of the value of $H_0$ that deviates by 3 to 5$\sigma$ between direct (late-time) measurements and indirect (or early-time) extrapolations, or the measurement of $S_8$, which shows a 2 to 3$\sigma$ tension between CMB and galaxy-lensing measurements. As shown during this conference, recent observations and  studies are providing new insights into these discrepancies, which might indicate a path towards reconciliation. In addition, the first-year results from DESI that were announced in April 2024 hint at an intriguing deviation from a simple \lcdm\ model, with  data that favor a description of dark energy by a time-evolving  equation of state.  DESI is a statistics-limited experiment and additional data  from the year-3 sample, which  are already collected, will soon provide even better insights into these exciting hints. With  the forthcoming data from additional and complementary Stage-IV dark energy surveys such as Euclid or Vera Rubin, the coming decade promises to be uniquely rich in new cosmology results. 

\vskip6pt




\end{document}